\documentstyle[aps,prc,twocolumn,floats,epsfig]{revtex}
\draft
\begin{document}

\renewcommand{\thefootnote}{\arabic{footnote}}

\twocolumn[\columnwidth\textwidth\csname@twocolumnfalse\endcsname

\title{Changes in $r$-process abundances at late times}

\author{R. Surman$^{1}$ and J. Engel$^{2}$}

\address{${}^{1}$Department of Physics, Union College, Schenectady, NY  12308} 

\address{${}^{2}$Department of Physics and Astronomy, CB3255,
         The University of North Carolina,
         Chapel Hill, North Carolina 27599-3255}

\date{March 15, 2001}

\maketitle

\addvspace{5mm}

%
%===========================================================================
%
\begin{abstract}
We explore changes in abundance patterns that occur late in the $r$ 
process.
As the neutrons available for capture begin to disappear, a quasiequilibrium
funnel shifts material into the large peaks at $A=130$ and $A=195$, and into
the rare-earth ``bump" at $A=160$.  A bit later, after the free-neutron
abundance has dropped and beta-decay has begun to compete seriously with
neutron capture, the peaks can widen.  The degree of widening depends largely
on neutron-capture rates near closed neutron shells and relatively close to
stability.  We identify particular nuclei the capture rates of which should
be examined experimentally, perhaps at a radioactive beam facility.

\end{abstract}

\addvspace{5mm}]

\narrowtext

\section{Introduction}
\label{s1}

The $r$ process, which synthesizes roughly half the elements with atomic
number $A>70$, proceeds through neutron capture and beta
decay \cite{Burbridge,Cowan,Meyerrev}.  Through most of the process, we know,
capture is much faster than beta decay, so that very neutron-rich and
unstable nuclei are temporarily created before disappearing as the process
peters out.  The question of where the nucleosynthesis occurs, however, is
still unanswered.  An evacuated bubble expanding behind a supernova shock
waves is a promising candidate for the site, but not yet the clear choice.
The scenario is more convincing if the initial expansion is very
rapid \cite{Qian}, but simple models of fast adiabatic expansion \cite{Frei}
imply that the neutron capture must finish in less than a second.  In
traditional simulations, where capture must often wait for a nucleus to beta
decay, the process takes 2 or 3 times that long.

Simulations from, e.g., ref.\ \cite{Frei} demonstrate that the entire process
can take place quickly if neutron capture populates nuclei farther from
stability (and thus shorter lived) than usually thought.  Despite initially
forming at lower $Z$ and $A$ than traditional work suggests, the simulated
abundance peaks end up at the right atomic numbers.  The apparent reason is
that nuclei in the peak move up quickly in $Z$ through an alternating
sequence of beta-decay and neutron capture near the end of the $r$ process,
when the supply of free neutron begins to run out but before neutron capture
completely stops.  But how do peaks maintain themselves during this late
time, as beta decay drives each nucleus at a different rate towards
stability?  This question is actually more general than the rapid-expansion
scenario; a quick move towards stability while the neutron abundance drops,
though most dramatic if the path is initially very far away, in fact
characterizes all bubble $r$-process simulations that produce something like
the correct abundance distribution.  And the question is linked to a broader
issue:  the significance of neutron capture once the process has slowed down
so that it must compete with beta decay.  What happens if the capture rates
then are faster or slower than we think?  In which nuclei do capture rates
have the largest effects on final abundances, and can the rates there be
measured?  These are the kinds of issues we address here.

The late stages of neutron capture and beta decay in the $r$ process are much
different from what precedes them.  When the neutron-to-nucleus ratio $R$ is
much larger than 1, equilibrium between neutron capture and
photodisintegration is a good approximation; the ``path" consisting of the
most abundant isotopes for each element $Z$ is far from stability and moves
relatively slowly.  The term ``steady state" is sometimes used to refer to
this period, which ends when the neutron-to-seed ratio $R$ falls below a few.
To see what happens next, we note that the neutron separation energies along
the path in $(n,\gamma) \leftrightarrow (\gamma,n)$ equilibrium are related
to the neutron number density $n_n$, which is proportional to $R$ for slowly
changing matter densities, by
\begin{equation} 
S_{n} \approx -kT\ln\left \{ {n_{n} \over 2} \left
({2\pi\hbar^2 \over m_n k T}\right )^{3/2} \right \}~ , 
\label{eq:pathloc}
\end{equation} 
where $m_n$ is the neutron mass.  When $R$ drops below about 1, a nucleus on
the path that beta-decays can no longer capture enough free neutrons to
return to the path so that the path itself must move instead, inwards to
higher neutron separation energy.  The increased average neutron binding
makes photodisintegration less effective, which in turn reduces the number of
free neutrons still further, in accordance with eq.\ \ref{eq:pathloc}.  The
process feeds on itself, causing $R$ to drop exponentially and the
$r$-process path to move quickly towards stability.  Soon $R$ becomes so
small that $(n,\gamma) \leftrightarrow (\gamma,n)$ equilibrium begins to
fail, and beta decay moves a good fraction of nuclei away from the path.
Eventually, beta decay becomes faster than neutron capture and all remnants
of $(n,\gamma) \leftrightarrow (\gamma,n)$ equilibrium vanish.  The inability
of equilibrium to maintain itself on the time scale of beta decay is usually
called ``freezeout".

To answer the questions about peak evolution and the significance of neutron
capture at late times, we focus on two competing effects.  The first, which
dominates just as $R$ falls below a few, when $(n,\gamma) \leftrightarrow
(\gamma,n)$ equilibrium still holds well, is a funneling of material into
moving peaks, most notably the small rare-earth peak \cite{Surman}, but also
the larger peaks at $A=130$ and 195.  As time passes, the funnel fights an
increasing tendency for the peaks to spread because beta decay and
beta-delayed neutron emission compete harder with neutron capture.  The
interplay of funneling and spreading will imply that uncertain neutron
capture rates, which are irrelevant as long as $(n,\gamma) \leftrightarrow
(\gamma,n)$ equilibrium holds, become important fairly near stability, and
should be determined there more precisely.

We support our contentions with simulations of the neutron-capture part of
the $r$ process.  In most, we explore late times without worrying about a
fast $r$ process.  We assume an exponential decay of temperature and density
with a relatively slow time scale of 2.8 seconds, the same as in ref.\
\cite{Meyer}.  The distribution of seed nuclei with which we start is the
post-alpha-process distribution of ref.\ \cite{Meyer}, and we vary the
initial temperature (and thus the initial density, since we keep the entropy
equal to 300) and neutron-to-seed ratio.  What we call our ``standard
simulation" is run under these conditions with the intial temperature fixed
at $T_{9}=1.5$.  In section \ref{s3}, we worry more about a fast $r$ process
and therefore use different conditions.  We do this in two ways:  the first
is to use use the same temperature and density dependence as described above,
but a greater initial density (and thus a lower entropy).  The faster neutron
capture at high density pushes the path farther from stability and leads to
the formation of the $A=195$ peak and the exhaustion of free neutrons in only
tenths of a second.  The second way is to use the conditions of Ref.\
\cite{Frei}, where the temperature and density drop with a time scale of 50
ms and then level off at low values ($T_9<1$, $\rho< 100~{\rm g/cm^3}$).  The
resulting drop in photodissociation rates again moves the path farther from
stability, and the $A=195$ peak forms in under a second.  All our simulations
use nuclear masses from ref.\ \cite{frdm} and beta-decay rates from ref.\
\cite{mol}.

The rest of this paper is organized as follows:  In Section \ref{s2}, we
discuss the action of funneling and spreading in the formation of the
rare-earth element bump.  Section \ref{s3} applies the same ideas to the
large peaks, with emphasis on the change in the peak's location and width as
the path moves.  The most important results appear in Section \ref{s4}, which
discusses neutron capture near the peaks and isolates particularly important
rates that should be measured.  Section \ref{s5} is a conclusion.

\section{Funneling and spreading in the rare earth region}
\label{s2}

Ref.\ \cite{Surman} presented the basic dynamics of the near-freezeout
funnel.  It concluded that for much of the time when $R<1$, the system is
still nearly in $(n,\gamma) \leftrightarrow (\gamma,n)$ equilibrium, even as
the path moves inwards.  A kink at $N=104,106$ soon develops (or grows
stronger) when the path approaches a deformation maximum, which acts like a
miniature closed shell.  The nuclei near the bottom of the kink are further
from stability than those at the top and thus have shorter beta-decay
lifetimes (see fig.\ 3 from Ref.\ \cite{Surman}, which shows the path
together with contours of constant beta-decay lifetime).  The rate at which
the path moves is governed by the average beta-decay lifetime along the path,
which typically corresponds to a nucleus in the kink.  The nuclei at the
below the kink have shorter lifetimes than this overall average, and tend to
beta-decay before the path moves, then capture neutrons in an attempt to stay
in equilibrium along the path.  The nuclei above the kink have longer
lifetimes and so do not usually beta-decay before the path moves, instead
photodissociating to keep up with a path that is moving away from them.  The
neutron abundance is so low at this time that neutrons are essentially
transferred from nuclei that photodisintegrate to those that capture.  The
net result is that nuclei both near the bottom and top of the kink funnel
into it as the path moves, leading to a peak in the final abundances, whether
or not one exists before $R=1$.

Another process, this one not discussed in Ref.\ \cite{Surman}, acts to
weaken the funnel.  As $R$ drops below 1, so does the rate at which neutrons
are captured, since it is proportional to the neutron abundance.  As a
result, a nucleus in or near the kink will not always have time to capture
neutrons after it has beta-decayed; it may first undergo another beta decay
and move away from the path of greatest abundances.  Nuclei can emit neutrons
following beta decay, moving them still further from the path.  Thus, part of
the growing peak begins to seep to lower neutron number $N$.  Together with
material from above the peak moving down in $N$, this process acts to wash
out the peak in both $N$ and $A$.  
%\footnote{Only this neutron emission (or capture)
%following beta decay alters the distribution in $A$.  Beta decay itself 
%decreases $Z$ but increases $N$, leaving $A$ unchanged.  If neutron emission 
%and capture were somehow to cease, beta decay would not affect the evolution 
%of abundance vs.\ $A$.}.

Funneling and spreading counter one another, but as noted in the
introduction, the two mechanisms reach their most effective points at the
different times.  Close to $R=1$, when the path begins its inward trek, there
are still enough neutrons so that spreading is slow and the funnel dominates.
At very late times, by contrast, $R$ is so small that $(n,\gamma)
\leftrightarrow (\gamma,n)$ equilibrium is seriously compromised and
spreading is substantial.  Eventually, neutron capture becomes slower than
beta decay and the system freezes entirely out of equilibrium.  After that
capture essentialy stops and the only thing affecting the abundance
distribution vs.\ $A$ is some final spreading from delayed neutron emission.
Beta decay without emission continues to move nuclei away from the
equilibrium path, altering the distribution in $N$, but has no effect on
abundances plotted vs.\ $A$ (as they usually are).

\begin{figure}[hbt]
\centerline{\epsfig{file=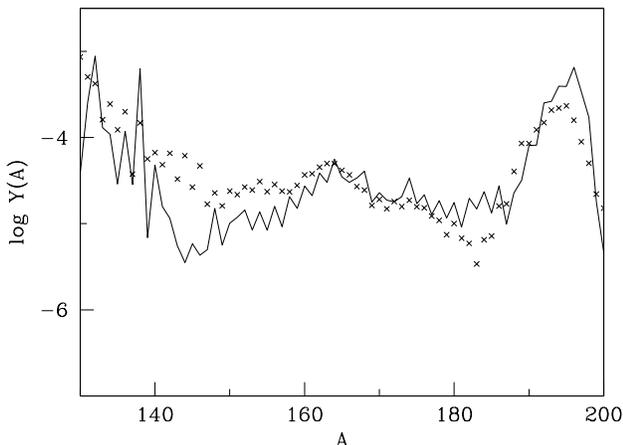,width=8.0cm}}
\caption{Predictions of our standard simulation (solid line) for the final 
abundances of $r$-process nuclides versus atomic number $A$, and the measured 
abundances (crosses) scaled to the simulation.  Note the peaks in the 
rare-earth element region near $A=160$.}
\label{ree1}
\end{figure}

\begin{figure}[hbt]
\centerline{\epsfig{file=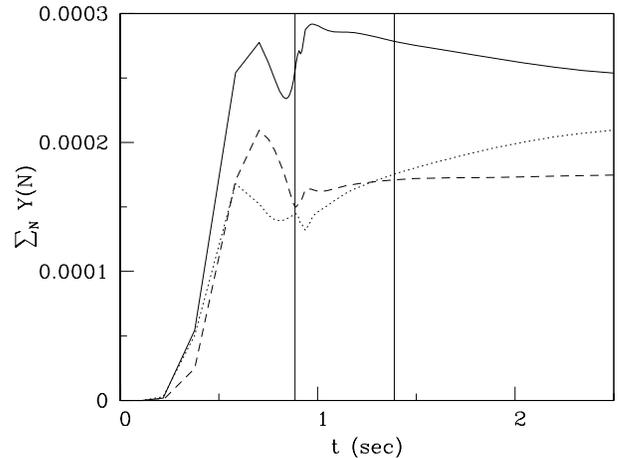,width=8.0cm}}
\caption{Results of our standard simulation for the total abundances in 
the regions just below the peak (dotted line), in the peak (solid line) and 
just above the peak (dashed line) as a function of time. A drop in peak 
material is suddenly reversed when the neutron/nucleus ratio $R$ nears 1,
a point indicated by the first solid vertical line.  The second solid
vertical line indicates the time of freezeout in the rare-earth region.}
\label{ree2}
\end{figure}

Our simulations make all these statements concrete.  Fig.\ \ref{ree1}
compares the results of our ``standard" simulation described in the
introduction with the measured abundances in the rare-earth region as a
function of $A$, showing the existence of a REE peak, in reality and in the
simulation.  The simulated peak clearly relies on a kink that develops in the
path because of the deformation maximum but, as argued in ref.\
\cite{Surman}, the mere existence of a kink is not sufficient to fully
produce such a peak; it achieves its full size only because of funneling.  To
see in more detail how the peak builds, we plot in fig.\ \ref{ree2} the
number of nuclei in three regions of $N$ --- that just below the peak ($N$=95
to 101), that including the peak ($N$=102 to 106), and that just above the
peak ($N$=107 to 113) --- as a function of time for the run just discussed.
The two vertical lines mark the points at which $R=1$, and at which
beta-decay and capture rates are equal in the rare-earth region, causing the
complete freezeout of $(n,\gamma) \leftrightarrow (\gamma,n)$ equilibrium
there.  The bump develops, and then actually starts to shrink as material
moves to higher $A$ early in the process.  But just before $R=1$, as the path
begins its inward move, it grows again.  As noted above, photodisintegration
is the dominant reaction above the bump and beta decay the important reaction
below, so that material on both sides of the bump shifts inwards.  After a
tenth or two of a second, spreading begins in earnest; nuclei in the bump
move slowly to lower $N$ and material moves down from above to fill the
trough above the peak, so that the abundance outside the peak starts to
increase.  To the right of the second vertical line, only beta decay
(sometimes with the emission of neutrons) occurs, so that material is shifted
downwards as the bump itself moves to lower $N$.

\begin{figure}[hbt]
\centerline{\epsfig{file=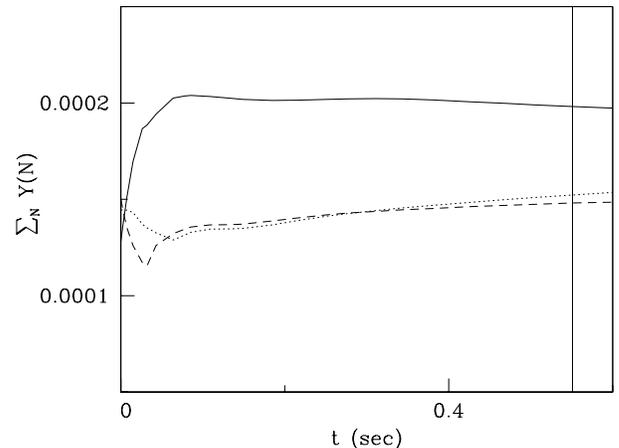,width=8.0cm}}
\caption{The same as fig.\ \protect\ref{ree2}, except that the abundances 
have been 
adjusted when $R=1$ so that the three regions all contain the same total.  
Now $R=1$ occurs at $t=0$ and the solid vertical line indicates the time
of freezeout. The formation of a peak here is due solely to funneling.}
\label{ree3}
\end{figure}

Although the brief region during which funneling dominates is evident in this
figure, one may wonder whether the peak could form even if it was absent
during the buildup phase along the path far from stability when $R >1$.  In
ref.\ \cite{Surman} we argued that that was the case, and fig.\ \ref{ree3}
here provides more evidence.  To make the figure we took the run discussed in
figs.\ \ref{ree1} and \ref{ree2} and adjusted the abundances at $R=1$ so that
the three regions of $N$ were equally populated.  We then let the run proceed
starting from $R=1$; the REE bump still formed.  Fig.\ \ref{ree3} clearly
shows that funneling in the first tenth of a second or so is responsible.
[Some material is brought in from outside the range of the plot.]  Here as
before, later times show effects of spreading and the post-freezeout shifting
of material downwards in $N$.

\section{Funneling and spreading in the formation of the $A=195$ peak}
\label{s3}

The fast $r$ process of Ref.\ \cite{Frei} relies on the formation of large
peaks farther from stability than usually thought.  Simulations show that the
fast process works but but do not explain how.  Why should a peak that forms
early at, e.g., $N=126$, remain there when the path moves as neutrons are
exhausted?  In more traditional simulations, when the path is assumed not to
move before freezeout, the usual explanation for peak buildup is approximate
``steady beta flow", which results in the longest live nuclides building up
the most.  But this kind of buildup takes as least as long as the lifetime of
the longest-lived nucleus, and the inward motion of the path we're discussing
here takes much less time.  Something like steady flow therefore cannot be
responsible for the existence of the peak at its final location in ref.\
\cite{Frei}.  What is?  The answer is a funneling phenomenon similar to that
we've already discussed, though slightly more complicated because instead of
a kink we now have a long ladder of $N=126$ isotopes populated at any given
time.

As already noted, the speed at which the path moves inward after $R<1$ is
given by the average rate of beta decay along the path.  This average rate
tends to occur along nuclei in the middle of the ladders at $N$ = 82 and 126.
Thus the nuclei below them decay, and then capture into the peak as long as
the funnel operates efficiently.  A decay followed by a capture moves a
nucleus up in $Z$, so material at the bottom of the ladder moves into the
center of the ladder as the funnel proceeds.  Further up the ladder, nuclei
neither capture nor photodissociate, since the path continues to run through
these nuclei even as it moves toward stability.  Instead they simply
beta-decay, but more slowly than nuclei at the bottom of the ladder.  The
result is that the entire ladder shortens as the bottom moves up faster than
the top.  Above the ladder, for $N$ just above 82 or 126, nuclei with slower
beta-decay rates photodissociate into the peak, adding material just as in
the REE region.  These dynamics combine to move the peak up slowly in $A$ at
the same time as they heighten and narrow it.  Though the large peaks clearly
form during the steady-state phase of the $r$ process, when $R >> 1$, they
are shaped and moved at later times as just described.

\begin{figure}[hbt]
\centerline{\epsfig{file=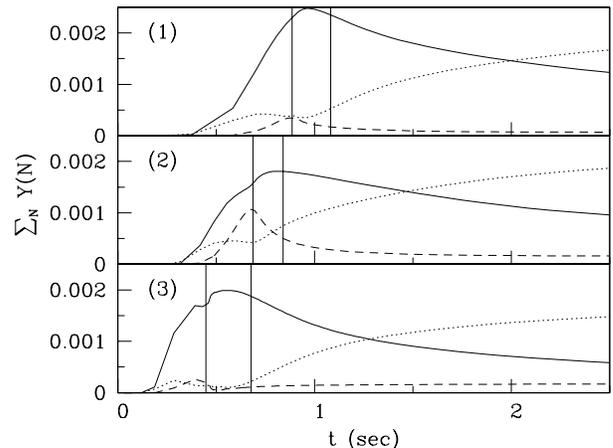,width=8.0cm}}
\caption{Total abundances in the regions just below the peak (dotted
line), in the peak (solid line), and just above the peak (dashed line) as 
a function of time for the three types of simulations discussed. Simulation 
(1) is the standard simulation, (2) uses the conditions of Ref.\  
\protect\cite{Frei}, and (3) is the standard simulation at much lower 
entropy.}
\label{195a} 
\end{figure}

\begin{figure}[hbt]
\centerline{\epsfig{file=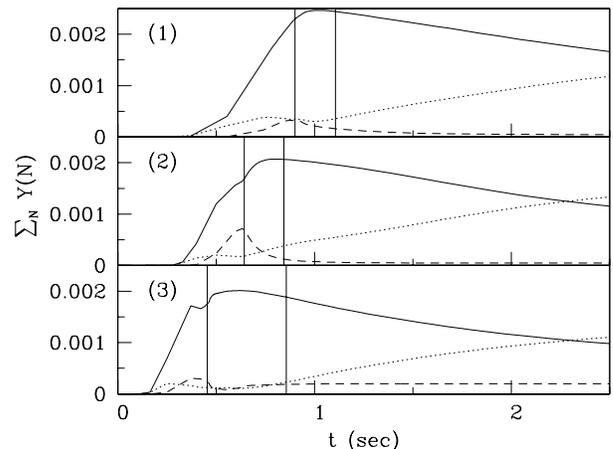,width=8.0cm}}
\caption{Same as fig.\ \protect\ref{195a}, but with simulations using a
newer set of neutron-capture rates from Ref.\ \protect\cite{rates3}.}
\label{195b} 
\end{figure}

These dynamics, however, can sometimes be masked by spreading.  Whether or
not they are depends on the temperature and density of the environment and on
nuclear properties.  Fig.\ \ref{195a} shows the funneling and spreading of
material around the $A=195$ peak in our standard simulation as well as the
two faster simulations described in the introduction.  As in fig.\
\ref{ree2}, we plot sums of abundances in three regions of $N$ --- below the
peak ($N=110-123$), within the peak ($N=124-126$), and above the peak
($N>126$) --- as a function of time.  In each case, the peak forms when
$R>1$, but continues to grow for the few tenths of a second following $R=1$.
Much of this material comes from the photodissociation of material above the
peak (dashed line in each plot).  At later times, spreading takes over as
material in the peak beta-decays to lower $N$.  The onset and rate of
spreading depends on how fast the neutrons are depleted after $R<1$, which in
turn depends on the temperature and density.  For a simulation at high
density, as in part (3) of fig.\ \ref{195a}, neutrons disappear very rapidly
when $R<1$, and so in a short time beta decay and beta-delayed neutron
emission win out completely over neutron capture.  When the density and
temperature drop quickly, as in part (2) of the figure, neutrons disappear
more slowly.  As a result, neutron capture competes with beta decay over a
longer period of time, delaying the full onset of spreading.  Nonevironmental
factors can also affect the balance between funneling and spreading.  In
anticipation of our discussion of neutron capture in Section 4, we run
simulations with the same three sets of conditions but a newer set of
calculated neutron-capture rates, from ref.\ \cite{rates3}.  Fig.\ \ref{195b}
shows the funneling and spreading of material in these simulations; the
latter is less effective than in fig.\ \ref{195a}.  We expand on this point
in the next section.

\begin{figure}[hbt]
\centerline{\epsfig{file=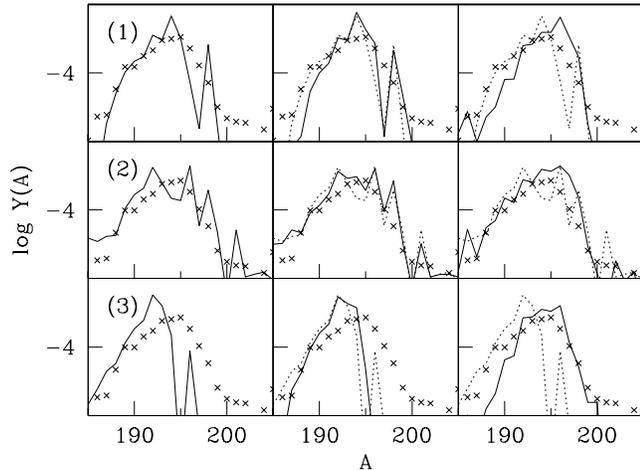,width=8.5cm}}
\caption{The evolution of the 195 peak in late times for the three types
of simulations, labeled as in fig.\ \protect\ref{195a}.  The first frame 
shows the
peak at $R\sim 1$, the third frame shows the final abundances, and the
second frame is taken from a time in between, when $R$ is less than 1 but
much larger than its value at freezeout.  For comparison, the dotted line
in the second and third frames replots the abundances from the first
frame. The scaled observed abundances are plotted as crosses.} 
\label{195c} 
\end{figure}

\begin{figure}[hbt]
\centerline{\epsfig{file=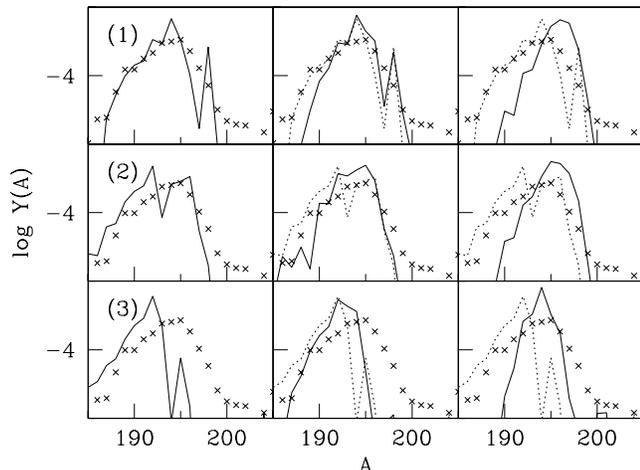,width=8.5cm}}
\caption{Same as fig.\ \protect\ref{195c}, but with simulations using a
newer set of neutron-capture rates from Ref.\ \protect\cite{rates3}.}
\label{195d}
\end{figure}

The way the peak moves and gets shaped can be seen in fig.\ \ref{195c}, which
shows its time development in the three types of simulations.  While the peak
initially narrows some as discussed above, it soon spreads, so that the
effect is barely visible.  Fig.\ \ref{195d} shows the same development but
with the newer capture rates.  Here the narrowing of the peaks is evident,
and is not erased by spreading at later times.  The shifting of the peak to
higher $A$ is apparent in both sets of plots, and is most pronounced in the
faster simulations where the peak forms much further from stability.

\section{Significance of capture rates}
\label{s4}

\begin{figure}[hbt]
\centerline{\epsfig{file=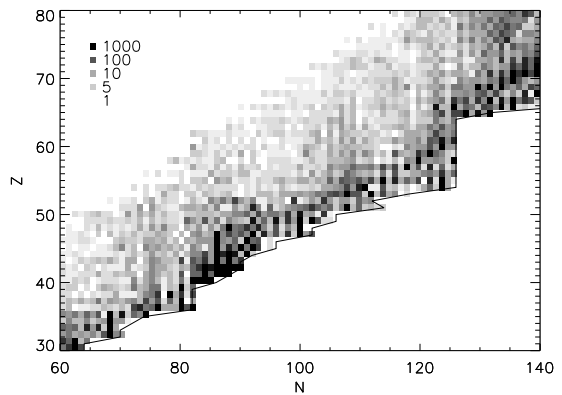,width=8.0cm}}
\caption{For each $N,Z$ the log of the ratio of the  highest neutron-capture 
rate in our set \protect\cite{Cowan,rates1,rates2,rates3} to the lowest.  The 
darkest squares correspond to ratios greater than 1000, as indicated in the 
key.}
\label{comprates}
\end{figure}

\begin{figure}[hbt]
\centerline{\epsfig{file=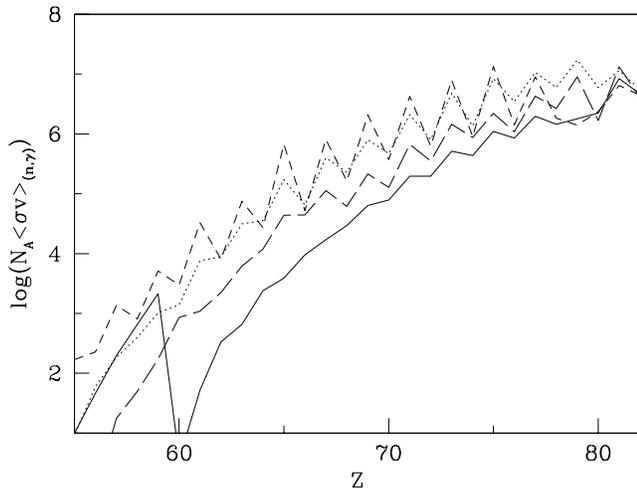,width=8.5cm}}
\caption{The four sets of neutron-capture rates, plotted as a function of 
$Z$, for $N=124$ just below the closed neutron shell.}
\label{195rates}
\end{figure}

\begin{figure}[hbt]
\centerline{\epsfig{file=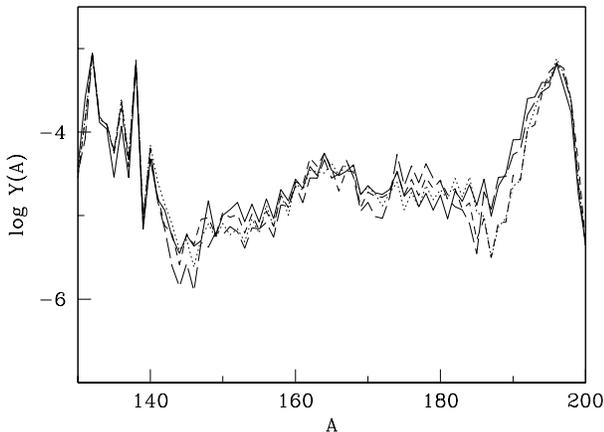,width=8.0cm}}
\caption{Abundance curves from our standard simulation with the 4 sets of 
neutron-capture rates.  Faster rates near the closed shell yield a narrower 
$A=195$ peak (note the left edge of that peak). }
\label{abunrates}
\end{figure}

Funneling operates unhindered just a short time before $(n,\gamma)
\leftrightarrow (\gamma,n)$ equilibrium falters and spreading sets in.  As we
saw in the last section, once spreading is important, neutron-capture rates
become so too.  They determine how likely a nucleus that has beta-decayed is
to return to the path before decaying again.  Fast rates mean that
$(n,\gamma) \leftrightarrow (\gamma,n)$ equilibrium hangs on longer and
spreading is delayed.  Thus, the ultimate degree of widening a peak
experiences depends on neutron-capture rates.  To illustrate this point, we
run simulations of the three types discussed above with four different sets
of calculated rates \cite{Cowan,rates1,rates2,rates3}.  These sets were
calculated with different models for nuclear masses, slightly different
treatments of the dominant statistical capture, and different assumptions
about the importance of direct capture.  Not surprisingly, the rates can
differ from one another significantly.  Fig.\ \ref{comprates} plots the ratio
of the smallest to largest rates as a function of $N$ and $Z$.  When we use
these rates in simulations (though all with the same mass model \cite{frdm})
we find variations in the final results for all values of $A$.  We continue
to focus on peaks, however, partly because the abundances are higher there
than in neighboring regions, so differences are more significant, and partly
because the differences in the left edge of the $A=195$ peak are particularly
noticeable.  As we already saw in the last section, and as figs.\
\ref{195rates} and \ref{abunrates} show in more detail, the peak doesn't
spread very much when rates are fast near the $N=126$ closed shell.  By
contrast the slowest rates at these points cause the widest final peaks.
These effects, incidentally, are particularly significant for the Ref.\
\cite{Frei} conditions, where $(n,\gamma) \leftrightarrow (\gamma,n)$
equilibrium falters earlier because of the rapid drop in temperature and
density, so that capture rates become important sooner.

\begin{figure}[hbt]
\centerline{\epsfig{file=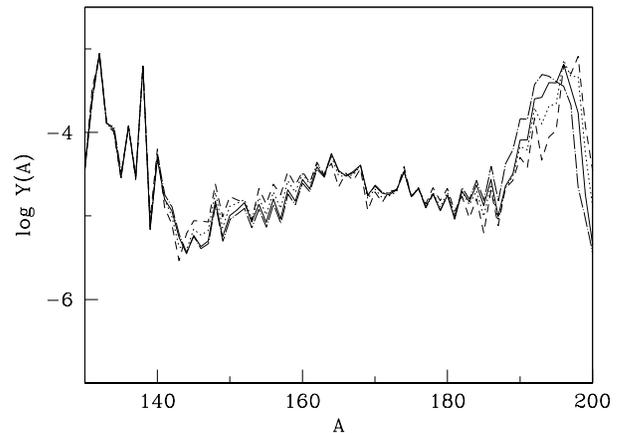,width=8.0cm}}
\caption{Abundance curve when the capture rates for $N=123-125$ alone are 
changed.  The solid line represents the results of with the rates from ref.\ 
\protect\cite{Cowan}, the dotted line the results when those rates are 
increased by 10, the dashed line the results when the rates are increased
by 100, and the dot-dashed line the results when the rates are shrunk by
100.}
\label{123125}
\end{figure}

We can see the role of capture near the peak even more clearly by changing
the rates only for $N$ between 123 and 125.  Fig.\ \ref{123125} shows the
results when those rates are multiplied by 10 or 100, or divided by 100.
When the rates increase, funneling becomes stronger and spreading weaker as
$(n,\gamma) \leftrightarrow (\gamma,n)$ equilibrium is partially restored.
As a result, the final abundance peak at $A=195$ narrows.

It so happens that the nuclei at and just below closed shells are notoriously
difficult to calculate \cite{goriely}.  Commonly used statistical methods may
not be applicable for all those nuclei because of the low density of states
at low energies \cite{goriely,rates3}.  Rates of direct capture, which also
plays a role, are uncertain because we don't know how much isovector dipole
strength lies low in nuclei far from stability.  To determine the
astrophysical parameters in the $r$-process environment, it's therefore
important to measure the rates in these nuclei where possible.  Of course
most of them are out of experimental reach for a long time to come.  But the
most important are actually relatively close to stability.

\begin{figure}[hbt]
\centerline{\epsfig{file=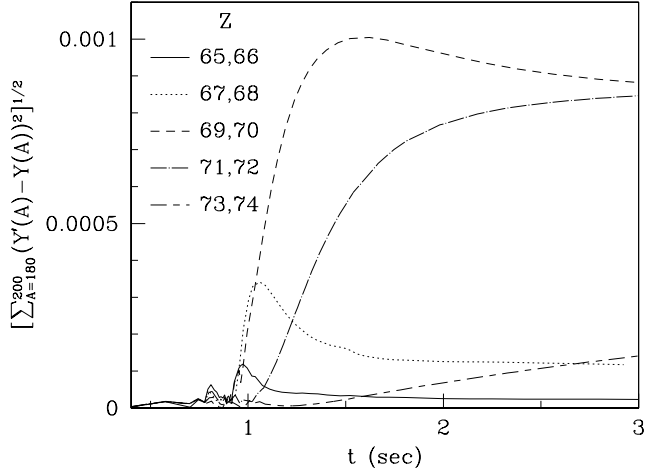,width=8.5cm}}
\caption{Root-mean-square differences between the abundances with the rates 
of \protect\cite{Cowan} and with those same rates everywhere except for a few 
nuclei with two values of $Z$ and $N=123-125$ (those are multiplied by 100), 
as a function of time.  We used the standard-simulation conditions. Each 
curve corresponds to increasing a different set of rates.  The nuclei with 
the largest effect on the final abundances have $Z=69-70$ and $71-72$.}
\label{select}
\end{figure}

\begin{figure}[hbt]
\centerline{\epsfig{file=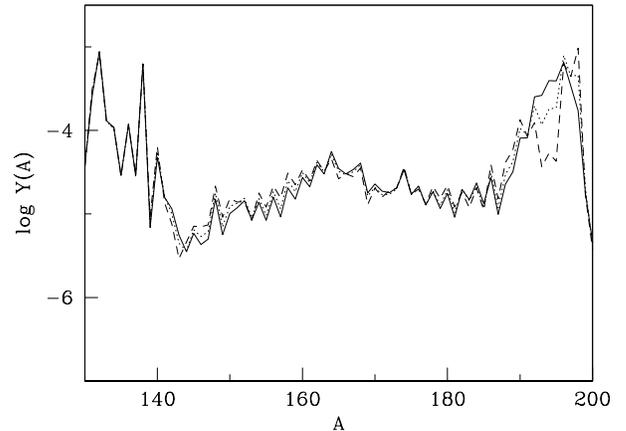,width=8.0cm}}
\caption{Final abundance curves in the simulation yielding fig.\
\protect\ref{select} when only the rates of nuclei with $N=123-125$ {\it and}
$Z=69-72$ are increased, by factors of 1 (solid line), 10 (dotted
line) and 100 (dashed line).  Note the results of these simulations are
nearly identical to those of fig.\ \protect\ref{123125}; the largest changes to
the final abundance distribution are due to the modification of the
capture rates of just these four nuclei.}
\label{selectres}
\end{figure}

To show why this is, we make even more selective changes, now for just 1 or 2
values of $Z$, in the rates for $N$ between 123 and 125.  Fig.\ \ref{select}
plots the root-mean-square difference between the abundance distribution
(within our standard simulation) when these rates are increased by 100 and
and when they are unaltered, as a function of time.  {\it The ultimate degree
of change depends strongly on which rates we change}.  Altering those below
$Z=69$ does little in the end because a) these nuclei are farther from
stability, where the system is closer to $(n,\gamma) \leftrightarrow
(\gamma,n)$ equilibrium and capture rates are nearly irrelevant, and b) Any
changes that do occur have time to be diluted by spreading.  Altering those
above $Z=72$ doesn't do much to the final abundance pattern because the
system has nearly frozen out of equilibrium, making neutron capture
irrelevant because there are so few free neutrons and the temperature has
dropped.  The nuclei for which changes do have large permanent effects lie
between $Z=69$ and 72 (Tm,Yb,Lu,and Hf), and correspond to the rough location
of the path just before full freezeout, when neutron capture and beta decay
compete on equal footing.  Then dramatic differences in flow result from
increasing the neutron-capture rates, differences that are not erased by
subsequent spreading.  We see similar effects when the rates are decreased
instead of increased.  Fig.\ \ref{selectres} shows final abundance curve for
a standard run with the rates of ref.\ \cite{Cowan} and standard runs in
which the capture rates of the nuclei with $Z=69-72$ and $N=123-125$ are
increased by 10 and 100.  The differences are significant.  All these
statements remain true both when we make wide variations in the initial
temperature, initial density, and time scale in our simulations, and when we
use the conditions of ref.\ \cite{Frei}.  The reason is that in this range of
$Z$, the calculated \cite{mol} beta-decay lifetimes of the $N=126$ nuclei
increase from about .07 s to 4.2 s\footnote{When we replace these lifetimes
with the more accurate and faster ones calculated in ref.\
\protect\cite{engel}, we find only slight differences in fig.\
\ref{select}.}.  The path there slows down around the peak, giving the
neutrons time to disappear through capture on nuclei in other regions.  This
happens whether in the time prior the path moves over a large distance
(quickly at first, because of the fast beta decay rates far from stability)
or over a shorter distance.  Although it's theoretically possible for
freezeout to occur for $Z < 69$, the resulting peak would almost certainly be
too low in $A$ to match the observed abundances.

The rates for these few nuclei are thus the ones on which the final
$r$-process abundances depend most sensitively, and measuring the associated
cross sections would be useful; we could then better constrain the
temperature and density during the $r$ process.  Unfortunately these nuclei
are still far enough from stability that their cross sections may not be
possible to measure, even partially through spectroscopic factors in transfer
reactions with radioactive beams at RIA.  A yield of about $10^4$/sec is
probably necessary for such experiments, while estimates \cite{ria} of
production at a RIA ISOL facility indicate that $Z$ must be about 77 before
yields will become that large.  But we can approach the nuclei we're
interested in, and see how measurements and calculations compare near the
most critical region.

\begin{figure}[htb]
\centerline{\epsfig{file=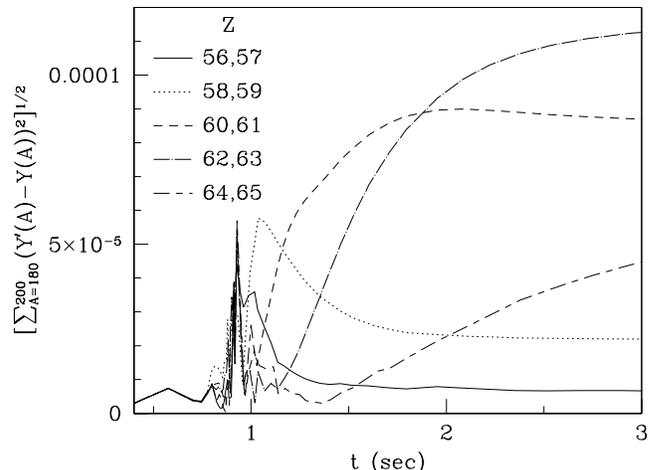,width=8.5cm}}
\caption{Same as fig.\ \protect\ref{select}, except for the rare-earth 
region, where 
rates of nuclei with $N=102-104$ and particular values of $Z$ are changed.  
The nuclei with the largest effect on the final abundances have $Z=62-63$.}
\label{reeselect}
\end{figure}

We're more fortunate in the rare-earth region because neutron-capture rates
are faster there than near the $A=195$ peak, and freezeout therefore occurs
closer to stability.  Fig.\ \ref{reeselect} shows what happens when we
selectively change the capture rates of nuclei just below the kink, with
$N=102-104$, for particular values of $Z$.  The nuclei with the strongest
effect now have $Z=62$ and 63 (Sm and Eu).  As before, the location of the
most important nuclei is not very sensitive to initial $r$-process
conditions.  These nuclei are actually within RIA's reach.  For the nucleus
$Z=62, N=102$ ($^{164}$Sm), yields should be about $10^5$/sec, for $Z=63,
N=102$ ($^{165}$Eu) they should be about $10^6$/sec, and for $Z=63, N=104$
($^{167}$Eu) about $10^4$/sec \cite{ria}.  Experiments to study their
capture cross sections are worth considering.

We have not discussed the $A=130$ peak in any detail.  Our conclusions there
are more limited because we do not reproduce the region below the peak very
well.  Abundances in that area are very sensitive to the outcome of the alpha
process, which we do not simulate.  Nonetheless we do get a peak at $A=130$
and generally find that varying the $N=79-81$ neutron-capture rates has the
largest effect for $Z=48-51$ ($^{127-129}$Cd, $^{128-130}$In, $^{129-131}$Sn,
$^{130-132}$Sb).  RIA should be able to make enough of these isotopes to
allow experiments.

\section{Conclusion}
\label{s5}

For most of the $r$ process, neutron-capture rates are irrelevant because
they are fast enough to maintain equilibrium with photodisintegration.  But
at late times, the situation is different.  Our investigation of funneling
and spreading led us to identify particular nuclei relatively close to
stability whose neutron-capture rates have significant effects on the shapes
of peaks.  It would be nice to have experimental information about these
nuclei, even if indirect, e.g.\ spectoscopic factors through $(d,p)$
neutron-transfer reactions.  Measurements of the capture rates themselves in
other nuclei closer to stability would also be useful; they would help tune
models, which could then be better extrapolated to the important nuclei
identified here.  A full understanding of the $r$ process would then be a
little closer.

We thank J.H.\ de Jesus, B.S.\ Meyer, M.  Smith, and P.  Tomasi for useful
discussions.  We were supported in part by the U.S.  Department of Energy
under grant DE--FG02--97ER41019.

%\newpage

\end{document}